# COMMUNICATION

### Evidence for coexistence of spin-glass and ferrimagnetic phases in BaFe$_{12}$O$_{19}$ due to basal plane freezing

Keshav Kumar,[a] Shrawan Kumar Mishra,[a] Ivan Baev,[b] Michael Martins,[b] and Dhananjai Pandey*[a]

**We present here results of low-temperature magnetization and x-ray magnetic circular dichroism studies on single crystals of BaFe$_{12}$O$_{19}$ which reveal for the first time emergence of a spin glass phase, in coexistence with the long-range ordered ferrimagnetic phase, due to the freezing of the basal plane spin component.**

Hexaferrites constitute an important family of compounds used in several technological applications such as permanent magnets in motors, credit cards, sonars, computer memories, spintronic devices, and microwave communications[1]. Recent years have witnessed revival of interest in these compounds following the discovery of type-II multiferroicity in the Y- and Z-type hexaferrites with strong magnetoelectric coupling around room temperature[2–4]. Further, the M-type hexaferrites have also evinced a lot of attention due to the discovery of several exotic quantum critical phenomena such as quantum paraelectricity (QPE)[5,6], quantum electric dipole liquid state (QEDL)[7,8], quantum tunneling of magnetization[9], quantum electric dipole glass[8], and magnetic quantum critical point[10]. Based on low-temperature dc magnetization (M(T)), ac susceptibility ($\chi(\omega, T)$) and x-ray absorption spectra (XAS)/x-ray magnetic circular dichroism (XMCD) studies on single crystals of BaFe$_{12}$O$_{19}$ (BFO), we report here another novel phenomenon resulting from freezing of the transverse (basal plane) component of the spins into a spin glass state at low temperatures. Our results show that BaFe$_{12}$O$_{19}$ falls in the category of the geometrically frustrated ordered compounds[11–16] showing exotic spin liquid, spin ice, and spin glass transitions even in the absence of any apparent substitutional disorder, with one very significant difference. Unlike the spin-glass phases in other geometrically frustrated compounds where it emerges from the high-temperature paramagnetic phase, the spin-glass phase of BFO emerges from the long-range ordered (LRO) ferrimagnetic (FIM) phase which continues to coexist with the spin-glass phase.

We have used flux-grown crystals of BFO in the present investigation. The details of crystal growth, characterization and physical property measurements are given in electronic supplementary information (ESI). The as-grown crystals are hexagonal platelet-shaped with well-developed facets as shown in Fig. 1(a). The crystallinity and symmetry of the crystals were checked using Laue diffraction pattern collected in the reflection geometry using a polychromatic beam incident along the c-axis of the hexagonal unit cell (see Fig. 1(b)). The presence of closely spaced diffraction spots along six symmetry-related directions not only confirms the crystallinity but also confirms the hexagonal symmetry of the as-grown crystals. The

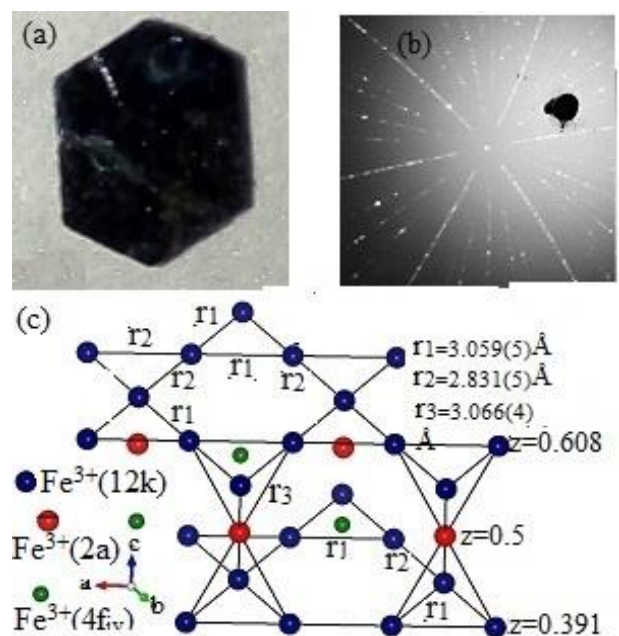

Fig. 1: (a) Photograph of as-grown crystal, (b) Laue pattern of BaFe$_{12}$O$_{19}$ single crystal with x-ray beam along [00l] direction, and (c) kagome bilayer configuration linked via pyrochlore slabs with different nearest neighbour bond lengths r$_1$, r$_2$, and r$_3$.

a. School of Materials Science and Technology, Indian Institute of Technology (Banaras Hindu University), Varanasi, India-221005.
b. Universität Hamburg, Institut für Experimentalphysik Luruper Chaussee 149, D-22761 Hamburg, Germany.

magnetoplumbite structure of the BFO in the P6₃/mmc space group was confirmed by Rietveld technique using x-ray powder diffraction pattern collected on calcined powder samples (see the ESI). The refined structural parameters and selected bond lengths are given in table S2 & S3 of ESI.

As per the classical Gorter model[17], the magnetic structure of barium hexaferrite comprises 3d⁵Fe³⁺ spins at the 2a, 2b, and 12k Wyckoff sites of the P6₃/mmc space group with spin up configuration and the spins at the 4f_IV and 4f_VI Wyckoff sites with spin down configuration, giving rise to an overall ferrimagnetic structure with a net magnetic moment of 20μ_B per formula unit[17]. This Ising like picture for the 3d⁵Fe³⁺ spins is, however, questionable, since a magnetic transition has been reported in the ab-plane with strong spin-phonon coupling[18]. The variation of $M_{\perp c}$(T) and $M_{//c}$(T), measured during warming cycle on a zero-field cooled (ZFC) crystal, with a magnetic field of 100 Oe applied perpendicular (⊥) and parallel (//) to the c-axis of the unit cell, respectively, shown in Fig. 2 reveals that $M_{\perp c}$(T) increases steadily with decreasing temperature upto ~40K and then shows a peak ~40K, whereas $M_{//c}$(T) decreases continuously with decreasing temperature. This confirms a magnetic transition at ~40K. Our results suggest that the spins are not fully aligned along the c-axis of BFO at low temperatures but have a significant component transverse to the c-axis in the basal plane (00l) due to the canting of the spins away from the c-axis.

In order to confirm the canting of the 3d⁵Fe³⁺ spins away from the c-axis, we investigated the angle-dependent XMCD signals using XAS spectra recorded on a single crystal of BFO at 30K in the normal and grazing incidence (GI) geometries where the angle (θ) between the direction of propagation vector of the circularly polarized soft x-ray beam and c-axis is 0⁰ and 15⁰, respectively. The incident-flux-normalized x-ray absorption spectra (XAS) obtained in the normal and GI geometries using left circularly and right circularly polarized x-ray photons corresponding to the Fe L₂, ₃ edges, labeled as σ₊ and σ₋, are shown in Figs. 3(a) and (b), respectively. The XMCD spectra (Δσ = σ₊ - σ₋) at iron L₂, ₃ edges for the normal and grazing angle incidence of polarized x-ray photon are shown in the same figure below the XAS spectra. From the angle-dependent XAS and XMCD spectra, spin magnetic moment can be calculated using the following spin sum rule equation[19,20].

$$m_{spin} + 7m_T^\theta = -\frac{(6P-4Q)\,n_h}{R} \qquad \ldots\ldots\ldots(1)$$

in which P = $\int_{L3}(\sigma_+ + \sigma_-)d\omega$, Q = $\int_{L3+L2}(\sigma_+ + \sigma_-)d\omega$], R = $\int_{L3+L2}(\sigma_+ + \sigma_-)d\omega$, $m_{spin}$ is the total spin magnetic moment in units of μ_B/formula unit, $n_h$ is the number of Fe 3d holes, $m_T^\theta = \langle T_\theta \rangle \mu_B/\hbar$ with $\langle T_\theta \rangle$ being the expectation value of magnetic dipole operator, and L₃ and L₂ represent the integration range over energies of the two absorption edges. Using equation (1), we obtained ($m_{spin} + 7m_T^{0^0}$)≈0.134μ_B/ion for the magnetic moment parallel to the c-axis and ($m_{spin}+7m_T^{15^0}$)≈0.06μ_B/ion for the transverse component of the moment. The observation of significant XMCD signal for the GI geometry clearly suggests that the 3d⁵Fe³⁺ spins are canted away from the c-axis. In order to further confirm the spin canting, we also analysed the XAS spectra and XMCD signals in the GI geometry recorded with dc field (H=100Oe) applied parallel to the beam direction (see

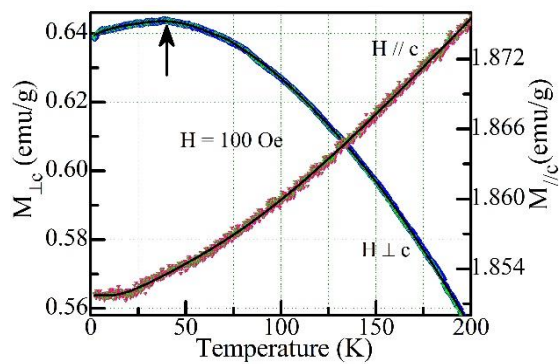

Fig. 2: Temperature evolution of the dc magnetization $M_{\perp c}$ and $M_{//c}$ of BaFe₁₂O₁₉ measured with dc bias field H =100 Oe applied perpendicular and parallel to the c-axis of unit cell, respectively.

Fig.3(c)). The significant enhancement of the XMCD signals in the presence of dc field further confirms that the 3d⁵Fe³⁺ spins are indeed canted away from the c-axis of BFO. Thus, both the dc magnetization M(T) and XMCD studies reveal that a finite component of the magnetic moments, which are primarily aligned parallel to the c-axis, lies in the ab-plane (i.e., basal plane (00l)) perpendicular to the c-axis.

In an isostructural compound SrCr₉ₓGa₁₂₋₉ₓO₁₉ (SCGO) with magnetoplumbite structure, it has been shown that the spins in the ab-plane undergo exotic spin liquid[21] and spin glass[22,23] transitions for Ga content 0<x ≤1 and 0.6≤x≤1[24], respectively, due to the formation of kagome bi-layer type block of spins in the basal plane. This geometry is fully frustrated[22] and leads to continuous macroscopic degeneracy[25,26]. The observation of significant component of spins in the ab-plane of BFO in M(T) and XMCD studies and occurrence of the magnetic transition around 40 K in $M_{\perp c}$(T) plots (see Fig.2) suggests that this transition may also have a spin glass character similar to SCGO. We therefore measured $M_{\perp c}$(T) during warming on zero-field cooled (ZFC) and field cooled (FC) single crystals and the typical results are depicted in Fig.4(a). The bifurcation of the ZFC and FC M(T) curves just above the peak temperature demonstrates the presence of history-dependent irreversibility characteristic of spin glass systems[27,28]. Since such an irreversibility may also arise due to blocking transition involving superparamagnetic (SPM) clusters[29], we analysed the variation of the real part of the ac susceptibility $\chi'_{\perp c}(\omega, T)$ with temperature (T), measured at various frequencies ($\omega=2\pi f$), shown in Fig.4(b). It is evident from this figure that the temperature $T_f(\omega)$ corresponding to the peak in the $\chi'_{\perp c}(\omega, T)$ curve shifts towards the higher side with increasing measuring frequency as observed in both the spin glass[27,28] and SPM system[29]. For SPM blocking transition, the temperature dependence of the spin relaxation time $\tau$ ($\simeq 1/\omega$) follows Arrhenius behaviour: $\tau = \tau_0 \exp(-\Delta E/k_B T)$, where $\tau_0$ is the attempt relaxation time, $\Delta E$ is the activation energy barrier and $k_B$ is the Boltzmann constant. For spin-glass systems, the spin dynamics, on the other hand, follows a power law behaviour[27,28]: $\tau = \tau_0((T_f - T_{SG})/T_{SG})^{-zv}$, where z is the exponent for power-law dependence of the correlation length $\xi$ on $\tau$ (i.e., $\xi \sim \tau^{-z}$) while v is the exponent for the temperature dependence of $\xi$ (i.e., $\xi \sim ((T_f - T_{SG})/T_{SG})^v$). For the SPM blocking transition, the $\ln(\tau)$ versus $1/T$ plot should be linear. The non-linear nature of the $\ln(\tau)$ versus $1/T$ plot shown in the inset of Fig.4(c) rules out the possibility of SPM blocking. On the other hand, least-squares fit to the $\ln(\tau)$ versus $\ln((T_f - T_{SG})/T_{SG})$ plot, as per the power-law behaviour using the procedure given in ref.[30], depicted in Fig.4(c) by continuous line, shows excellent fit. This confirms that the transition in the ab-plane of BFO is due to the critical slowing down of the spin dynamics expected for a spin glass transition. The best fit obtained between the observed (red circle) and calculated (black solid line) relaxation time ($\tau$) in Fig.4(c) using the power-law dynamics corresponds to $\tau_0 = (8.2 \pm 0.8) \times 10^{-5}$ s, $zv = (0.50 \pm 0.02)$ and $T_{SG} = (46.035 \pm 0.005)$K. The large value of $\tau_0$ ($\tau_0 \sim 10^{-5}$ s) suggests cluster spin glass state in BFO, since the characteristic time ($\tau_0$) required to flip a single magnetic spin in atomic glasses is $\sim 10^{-13}$s[28,31]. The calculated $zv$ value for BFO is anomalously small as compared to conventional spin-glass systems where it lies in the range range 4 to 12[28]. However, several oxide and alloy-based systems have been reported to exhibit low $zv$ values in the range $2 < zv < 4$[32], $1 \le zv < 2$[32,33] and $zv < 1$ ($zv = 0.9, \& 0.55$)[34,35]. Such spin-glass systems have been termed as unconventional spin-glasses[34]. The existence of spin-glass state at low temperatures in the ab-plane was further confirmed by studying the relaxation of thermoremanent magnetization (TRM) $M_{\perp c}(t)$ as a function of time[13,36,37]. For this, the sample was cooled from 300K to 40K and then held at this temperature for 600 sec in the presence of a dc magnetic field of 1000 Oe applied perpendicular to the c-axis. After the elapse of 600 seconds, the magnetic field was switched off and the decay of magnetization $M_{\perp c}(t)$ as a function of time was recorded for 8 hours. Fig.4(d) depicts the time evolution of $M_{\perp c}(t)$. The decay of TRM in spin glass systems is known to follow Kohlrausch-Williams-Watt (KWW) type stretched exponential behaviour[13,36,37]: $M(t) = M_0 + M_g \exp\{-(t/\tau_r)^\beta\}$, where $M_0$ is the ferromagnetic component due to the coexistence of ferrimagnetic phase, $M_g$ is the glassy component and $\tau_r$ is the characteristic relaxation time. For $\beta = 1$, the system follows exponential decay: $M(t) = M_0 + M_g \exp\{-t/\tau_r\}$ with a single relaxation time ($\tau$) corresponding to the potential energy barriers of equal height. Since the energy landscape for spin glass systems is multivalleyed with different barrier heights, the relaxation time ($\tau$) has a distribution[27,28] and $\beta$ lies in the range $0 < \beta < 1$[13,27,28,36-38]. A least-square fit using the KWW function gives an excellent match between the observed and fitted $M_{\perp c}(t)$ for $M_0 = 0.115 \pm 0.0001$, $M_g = 0.015 \pm 0.00018$, and $\beta = 0.63 \pm 0.01$. The value of $\beta = 0.63$ rules out the possibility of relaxation of TRM due to the pure long-range ordered ferrimagnetic phase of BFO and confirms the presence of metastable states due to the glassy phase.

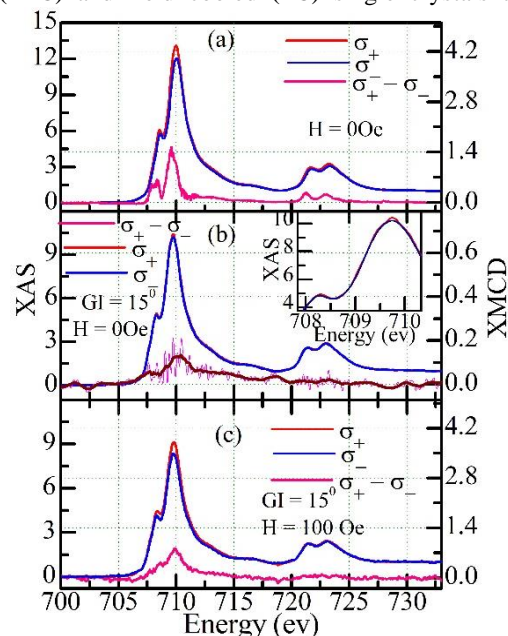

Fig. 3: X-ray absorption spectra (XAS) and x-ray magnetic circular dichroism (XMCD) signal at Fe L₂,₃-edges of BaFe₁₂O₁₉ recorded in (a) normal incidence geometry, (b) and (c) grazing angle geometry.

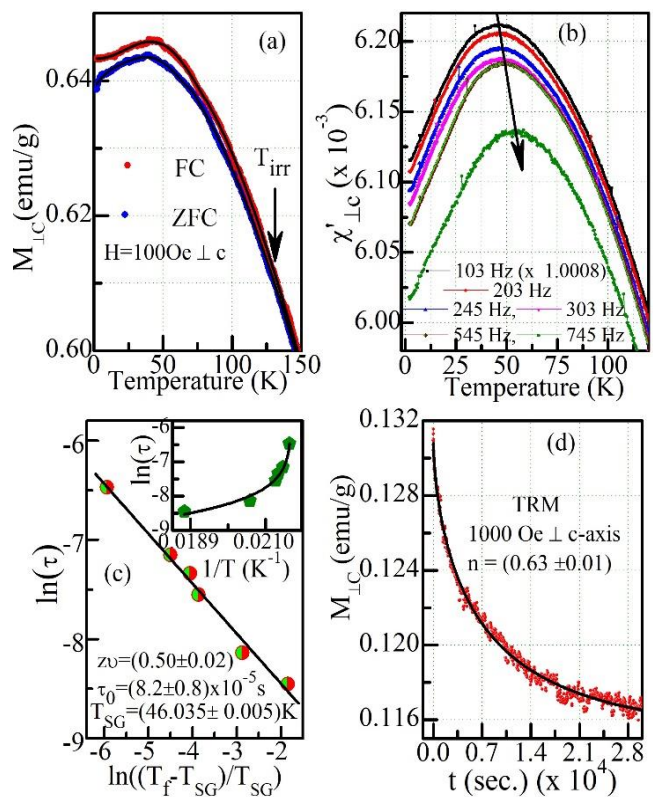

Fig.4: (a) Temperature dependence of ZFC & FC magnetization, (b) Evolution of real part of ac susceptibility measured at various frequencies (c) shows the power law fittings; inset shows the non-linear behaviour of the relaxation time and (d) Evolution of thermoremanent magnetization with time at 40K.

The spin-glass transition has conventionally been linked with substitutional site-disorder that is responsible for frustration and randomness of the interactions[27,28,39]. In this context, the observation of spin-glass transition in an ordered compound like BFO without any substitutional disorder may appear somewhat intriguing. However, spin-glass phase has been reported in several geometrically frustrated ordered compounds like pyrochlores (e.g., $Tb_2Mo_2O_7$[14]), hydronium jarosites (e.g., $(H_3O)Fe_3(SO_4)_2(OH)_6$[12]) and spinels (e.g., $MAl_2O_4$[40] with M=Co, Fe, and Mn). Recent theoretical studies on spin-glass transition in geometrically frustrated ordered compounds suggest that the presence of an infinitesimal disorder due to anisotropic exchange interactions or magnetoelastic strains is enough to stabilize the spin-glass state[41,42]. The spin-glass phase reported in such geometrically frustrated compounds including BFO is different from those reported in CaBaFe$_4$O$_{7+\delta}$[43], GaFeO$_3$, Mn3In and Na$_2$Mn$_3$(SO$_4$)$_3$($\mu_3$-OH)$_2$($\mu_2$-OH$_2$)$_2$ like compounds where there is inherent cation disorder due either to mixed-valence states or anti-site disorder or randomness in the partial occupancy of the Wyckoff position. BFO is isostructural with SCGO where a kagome bi-layer configuration of the spins has been reported for the basal plane spins. We depict in Fig.1(c) the possible kagome bi-layer configuration of spins with nearest neighbour bond-lengths for BFO in analogy with SCGO. We believe that the formation of kagome bi-layer configuration for the basal plane component of the spins in BFO along with the anisotropic exchange interactions due to variation in the nearest neighbour distances between 3d$^5$Fe$^{3+}$ spins parallel and perpendicular to the c-axis is responsible for the stabilization of the spin-glass phase observed by us in BFO. However, the analogy with SCGO and other geometrically frustrated ordered compounds like Tb$_2$Mo$_2$O$_7$[14] and (H$_3$O)Fe$_3$(SO$_4$)$_2$(OH)$_6$[12] ends here. The spin-glass phase in all other geometrically frustrated ordered compounds emanates from the paramagnetic phase whereas the spin-glass phase of BFO results from freezing of the basal plane component of the 3d$^5$Fe$^{3+}$ spins in the LRO ferrimagnetic phase which continues to coexist with the spin-glass phase. In this context, this is the first report of coexistence of spin glass and LRO phases in an ordered compound in the absence of any substitutional disorder. We believe that the present results would stimulate further theoretical and experimental studies to understand the origin of spin glass transition due to emergence of geometrical frustration in LRO magnetic systems.

**Acknowledgments:** Portions of this research were carried out at the light source PETRA III of DESY, a member of the Helmholtz Association (HGF). Financial support from the Department of Science and Technology (Government of India) within the framework of the India@DESY is gratefully acknowledged. Author K. Kumar is thankful to Dr. Arun Kumar for his help.

**Conflicts of interest:**
There are no conflicts to declare.